\documentclass[aps,prb,twocolumn,superscriptaddressgroupaddress,floatfix]{revtex4-2}
\usepackage{amssymb}
\usepackage{graphicx}
\usepackage{times}
\usepackage{amsmath}
\usepackage{amsthm}
\usepackage{amsfonts}
\usepackage[T1]{fontenc}
\usepackage[latin9]{inputenc}
\usepackage{array}
\usepackage{multirow}
\usepackage{color}
\usepackage{bm}
\usepackage{color}

\usepackage{bbm}
\usepackage{hyperref}
\usepackage{babel}
\usepackage[mathscr]{euscript}
\usepackage{float}
\usepackage{hyperref}
\hypersetup
{	colorlinks,%
	citecolor=green,%
	linkcolor=red,%
	urlcolor=blue%
}
\allowdisplaybreaks[4]

\begin{document}
\title{Wiedemann-Franz law in graphene in the presence of a weak magnetic field}
\author{Yi-Ting Tu}
\author{Sankar Das Sarma}
\affiliation{Condensed Matter Theory Center and Joint Quantum Institute, Department of Physics, University of Maryland, College Park, Maryland 20742, USA}

\begin{abstract}
  The experimental work [J. Crossno {\it et al.}, \href{https://doi.org/10.1126/science.aad0343}{Science {\bf 351}, 1058 (2016)}], which reported the violation of the Wiedemann-Franz law in monolayer graphene characterized by a sharp peak of the Lorenz ratio at a finite temperature, has not been fully explained.
  Our previous work [Y.-T.~Tu and S.~Das Sarma, \href{https://doi.org/10.1103/PhysRevB.107.085401}{Phys.\ Rev.\ B {\bf 107}, 085401 (2023)}] provided a possible explanation through a Boltzmann-transport model with bipolar diffusion and an energy gap possibly induced by the substrate. In this paper, we extend our calculation to include a weak magnetic field perpendicular to the graphene layer, which is experimentally relevant, and may shed light on the possible violation or not of the Wiedemann-Franz law.
  We find that the magnetic field enhances the size of the peak of the Lorenz ratio but has little effect on its position, and that the transverse component of the Lorenz ratio can be either positive or negative depending on the parameter regime.
  In addition, we do the same calculation for bilayer graphene in the presence of a magnetic field and show the qualitative similarity with monolayer graphene.
Our work should motivate magnetic-field-dependent experiments elucidating the nature of the charge carriers in graphene layers.
\end{abstract}

\maketitle

\section{Introduction}
In our previous work~\cite{Tu2023}, we proposed a simple theory that qualitatively explains the apparent violation of the Wiedemann-Franz (WF) law in monolayer graphene (MLG) reported in the experimental work of Crossno {\it et al.}~\cite{Crossno2016}.
The WF law states that the Lorenz number $L=\kappa/(\sigma T)$ is a universal constant in metals, $L_0=\frac{\pi^2}{3}\left(\frac{k_B}{e}\right)^2$~\cite{Franz1853,Lorenz1881}. Here, $\kappa$ and $\sigma$ are the thermal and electrical conductivities of the charge carrier, $k_B$ and $e$ are the Boltzmann constant and the electron charge, respectively.
This law is largely satisfied by normal metals described by Fermi liquids, but can sometimes be violated due to inelastic scattering effects~\cite{Mahajan2013,Lavasani2019,Ahn2022} or the bipolar diffusion effect~\cite{Tu2023}.
In Ref.~\cite{Crossno2016}, the authors reported a large violation of the WF law in MLG characterized by a high peak of $L/L_0\sim 20$ at a finite temperature $T\sim 60\,\mathrm{K}$, and attributed this violation to the ``non-Fermi liquid'' hydrodynamic effect of the quantum Dirac fluid nature of intrinsic graphene.
However, even with six free parameters, the finite temperature peak cannot be well-explained with the hydrodynamic theory~\cite{Lucas2016}, and that experimental observation in Ref.~\cite{Crossno2016} remains unexplained.

In Ref.~\cite{Tu2023}, we proposed an alternative but much simpler theory based on Boltzmann transport theory, where the scattering by short- and long-range impurities and acoustic phonons are treated phenomenologically.
No Fermi liquid violations or exotic interaction effects were included in our conventional Boltzmann theory in Ref.~\cite{Tu2023}.
We demonstrated that the bipolar diffusion effect, arising from the thermally induced electrons and holes around the Dirac point, produces a finite-temperature peak of $L/L_0$, which can be very high if we assume an (unintentional and uncontrolled) energy gap opening at the Dirac point, which is possible in experiments due to the presence of the hBN substrate underlying the graphene layers~\cite{DeanPrivate,Hunt2013,RibeiroPalau2018,Finney2019,Yankowitz2018}.
However, we do not claim that this theory unambiguously explains all of the experimental observations, and in particular, the presence of a gap at the Dirac point must be validated for our theory to explain the finite temperature peak reported in Ref.~\cite{Crossno2016}.
Indeed, it is unknown whether the hBN substrate really induces such a gap, and if so, what the size of the gap and the shape of the band near the gap could be.
More experimental work is needed to settle down the best explanation of that observation.
In particular, it is important that the findings of Ref.~\cite{Crossno2016} are reproduced or revised experimentally with more data so that we have a more complete picture of the situation.

One way to extend the original experiment~\cite{Crossno2016} is to add a magnetic field perpendicular to the graphene surface~\cite{KimPrivate}.
In this way, many qualitative behaviors can be checked with the theory.
Will the finite-temperature peak of $L/L_0$ be enhanced or suppressed by the magnetic field?
Will the position of the peak shift?
In addition, the magnetic field creates transverse motions of the electrons and holes, which give more complex features such as the possible change in the sign of the transverse component of the Lorenz ratio.
But, adding a magnetic field also complicates the physics because now there are three independent parameters controlling the system: temperature, doping, and magnetic field. In addition, there could be the fourth additional uncontrolled parameter associated with the energy gap.

This paper is the follow-up to and extension of Ref.~\cite{Tu2023}. We include a weak magnetic field in the Boltzmann transport theory of Ref.~\cite{Tu2023}, while keeping everything else the same.
We find that the size of the finite-temperature peak of $L/L_0$ is enhanced by the magnetic field, while the position does not change much. In addition, the sign of the transverse component can either be positive or negative depending on the parameter regime.
In addition, we repeat the same calculation for bilayer graphene (BLG), which is also experimentally relevant, and find similar qualitative results.
These observations can be tested experimentally to provide a step towards explaining the intriguing observations in Ref.~\cite{Crossno2016}.

The rest of this paper is organized as follows: 
In Sec.~\ref{sec:theory}, we present the setup of our theory, that is, the Boltzmann transport theory with magnetic field and bipolar diffusion included. 
In Secs.~\ref{sec:MLG} and \ref{sec:BLG}, we present the models and results for MLG and BLG, respectively. We conclude this paper in Sec.~\ref{sec:conclusion}.

\section{Theory}\label{sec:theory}

Our starting point is the Boltzmann equation
\begin{equation}
  \frac{\partial f}{\partial t} + \dot{\mathbf{r}}\cdot \frac{\partial f}{\partial \mathbf{r}} + \dot{\mathbf{k}}\cdot \frac{\partial f}{\partial \mathbf{k}} = \mathcal{I}\{f\}\,,
\end{equation}
where $f(\mathbf{r},\mathbf{k},t)$ is the distribution of electron wave packets at position $\mathbf{r}$, wavevector $\mathbf{k}$, and time $t$.
The semiclassical equations of motion are
\begin{align}
  \dot{\mathbf{r}}&=\frac{1}{\hbar}\frac{\partial\varepsilon}{\partial\mathbf{k}}-\dot{\mathbf{k}}\times\mathbf{\Omega}\label{eq:rdot}\\
  \dot{\mathbf{k}}&=-\frac{e}{\hbar}(\mathbf{E}+\dot{\mathbf{r}}\times\mathbf{B})\,,
\end{align}
where $\mathbf{E}$ ($\mathbf{B}$) is the applied electric (magnetic) field and $\mathbf{\Omega}$ is the Berry curvature.
We assume that $f$ is time-independent (steady state) and $f=f_0+\delta f$ for a small perturbation $\delta f$ near the local equilibrium
\begin{equation}
  f_0(\mathbf{r},\mathbf{k}) = \frac{1}{\exp{\frac{\varepsilon(\mathbf{k})-\mu(\mathbf{r})}{T(\mathbf{r})}}+1},
\end{equation}
In this paper, we measure energies directly in the units of temperature (Kelvins), so Boltzmann's constant $k_B$ equals unity in the formulas.
In the case that the magnetic field is weak in the sense that the cyclotron radius is much larger than the Fermi wavelength, we can neglect Landau quantization as well as the effect of the Berry curvature.
We restrict our theory entirely to the weak field semiclassical regime so that the magnetic field only adds a transverse force on the carriers without affecting anything else.
In the linear response regime of the applied electrochemical force $\mathbf{\mathcal{E}}=E+\frac{1}{e}\nabla\mu$ and temperature gradient $\nabla T$, the Boltzmann equation can be linearized as~\cite{Blatt1976}
\begin{equation}
  \mathbf{v}\cdot\left(e\mathbf{\mathscr{E}}+\frac{\varepsilon-\mu}{T}\nabla T\right)\left(-\frac{\partial f_0}{\partial\varepsilon}\right)-\frac{e}{\hbar}\mathbf{v}\times\mathbf{B}\cdot\frac{\partial\delta f}{\partial\mathbf{k}} = -\frac{\delta{f}}{\tau},
\end{equation}
where $\mathbf{v}=\frac{1}{\hbar}\frac{\partial\varepsilon}{\partial\mathbf{k}}$ is the velocity ($\mathbf{v}=v\hat{\mathbf{k}}$ for a scalar function $v(k)$ in our case), and we have used the relaxation time approximation for the collision term with relaxation time $\tau$, which may depend on $\varepsilon$ as well as $T$.

We only consider the case where $\mathbf{B}=B_z \hat{\mathbf{z}}$ is perpendicular to the surface of the material.
Since our system is rotationally symmetric, without any loss of generality, the transport coefficients can be calculated by assuming $\mathbf{\mathscr{E}}$ and $\nabla T$ to be in the $\hat{\mathbf{x}}$ direction, and then solving the differential equation for $\delta f$, and by plugging into the expressions for electrical and thermal currents:
\begin{align}
  J_{x,y} &= -eg_sg_v\int\frac{d^2k}{(2\pi)^2} v_{x,y} \delta f\nonumber\\
          &= \mathscr{E}_x(L_{EE})_{xx,yx} + \nabla T_x(L_{TE})_{xx,yx}\nonumber\\
  Q_{x,y} &= g_sg_v\int\frac{d^2k}{(2\pi)^2} v_{x,y} (\varepsilon-\mu) \delta f \nonumber\\
          &= \mathscr{E}_x(L_{ET})_{xx,yx} + \nabla T_x(L_{TT})_{xx,yx}
\end{align}
to extract the coefficients (here we restrict ourselves to a single band, and the degeneracies are $g_v=g_s=2$ in our case). The resulting formulas are~\footnote{The factor of $\frac{1}{2}$ was missing in Ref.~\cite{Tu2023}, but is irrelevant in the calculation of the Lorenz ratio.}
\begin{align}
  (L_{EE})_{xx} &= \frac{e^2}{2}\int d\varepsilon \left(-\frac{\partial f_0}{\partial\varepsilon}\right)D \frac{\tau v^2}{1+\left(\frac{e\tau v B_z}{\hbar k}\right)^2}\nonumber\\
  (L_{EE})_{yx} &= \frac{e^2}{2}\int d\varepsilon \left(-\frac{\partial f_0}{\partial\varepsilon}\right)D \frac{\left(\frac{e\tau v B_z}{\hbar k}\right)\tau v^2}{1+\left(\frac{e\tau v B_z}{\hbar k}\right)^2}\nonumber\\
  (L_{TE})_{xx} &= -\frac{e}{2}\int d\varepsilon \left(-\frac{\partial f_0}{\partial\varepsilon}\right)D \frac{\tau v^2}{1+\left(\frac{e\tau v B_z}{\hbar k}\right)^2}(\varepsilon-\mu)\nonumber\\
  (L_{TE})_{yx} &= -\frac{e}{2}\int d\varepsilon \left(-\frac{\partial f_0}{\partial\varepsilon}\right)D \frac{\left(\frac{e\tau v B_z}{\hbar k}\right)\tau v^2}{1+\left(\frac{e\tau v B_z}{\hbar k}\right)^2}(\varepsilon-\mu)\nonumber\\
  (L_{TT})_{xx} &= -\frac{1}{2T}\int d\varepsilon \left(-\frac{\partial f_0}{\partial\varepsilon}\right)D \frac{\tau v^2}{1+\left(\frac{e\tau v B_z}{\hbar k}\right)^2}(\varepsilon-\mu)^2\nonumber\\
  (L_{TT})_{yx} &= -\frac{1}{2T}\int d\varepsilon \left(-\frac{\partial f_0}{\partial\varepsilon}\right)D \frac{\left(\frac{e\tau v B_z}{\hbar k}\right)\tau v^2}{1+\left(\frac{e\tau v B_z}{\hbar k}\right)^2}(\varepsilon-\mu)^2\label{eq:transport}
\end{align}
and that $L_{ET}=-\frac{1}{T}L_{TE}$, where $D$ is the density of states.
The set of equations defined by Eq.~(\ref{eq:transport}) are the finite-magnetic-field generalization of the basic Boltzmann transport theory for our problem.

Now the above can be calculated for each band, and the total transport coefficients are the sums of them (we do not consider interband scatterings here).
The calculation is done with fixed carrier density $n$, where the chemical potential $\mu$ is obtained self-consistently
\begin{equation}\label{eq:consistent}
  n=\int_{+} d\varepsilon D_+f_0-\int_{-} d\varepsilon D_-(1-f_0)\,.
\end{equation}
where the range of the first (second) integral is in the conduction (valance) band, and $\pm$ denote the band indices.

Now the electrical and thermal conductivity matrices are
\begin{align}
  \sigma &= L_{EE} \nonumber\\
  \kappa &= L_{TT} - L_{TE}L_{EE}^{-1}L_{ET}.
\end{align}
Here, the bipolar diffusion effect is automatically included.
We define the effective Lorenz number componentwise
\begin{equation}
  L_{xx} = \frac{\kappa_{xx}}{\sigma_{xx}T},\quad L_{xy} = \frac{\kappa_{xy}}{\sigma_{xy}T}.
\end{equation}
For both components, the Lorenz number equals $L_0=\frac{\pi^2}{3e^2}$ in the regime where the Wiedemann-Franz law is satisfied, so below we will present the results using the Lorenz ratio $L_{xx,xy}/L_0$.
Note that such a componentwise treatment for the Lorenz ratio in the presence of a magnetic field was used also in hydrodynamic theory~\cite{Lucas2018}. 

\section{Monolayer graphene}\label{sec:MLG}

\begin{figure*}
  \includegraphics[trim=0 0 0 0, clip]{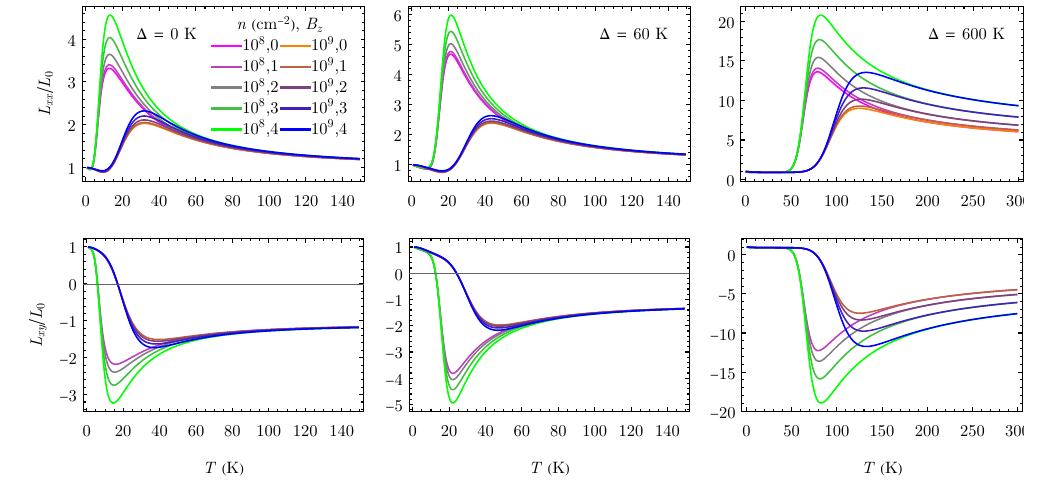}
  \caption{
    Longitudinal and transverse Lorenz ratio for MLG as a function of $T$, for various choices of $\Delta$, $n$, and $B_z$.
    The scattering parameters in the cases of $\Delta=0\,\mathrm{K}$ (first column) and $\Delta=60\,\mathrm{K}$ (second column) are $A/C=2.78\times10^{-5}\,\mathrm{K}^{-2}$, $B/C=4.63\times10^{-6}\,\mathrm{K}^{-3}$, and that in the case of $\Delta=600\,\mathrm{K}$ (third column) are $A/C=2.78\times10^{-9}\,\mathrm{K}^{-2}$, $B/C=4.63\times10^{-12}\,\mathrm{K}^{-3}$.
    The unit of $B_z$ is $0.2\,Ce^{-1}v_F^{-2}$.
  }
  \label{fig:MLG_T}
\end{figure*}

\begin{figure*}
  \includegraphics[trim=0 0 0 0, clip]{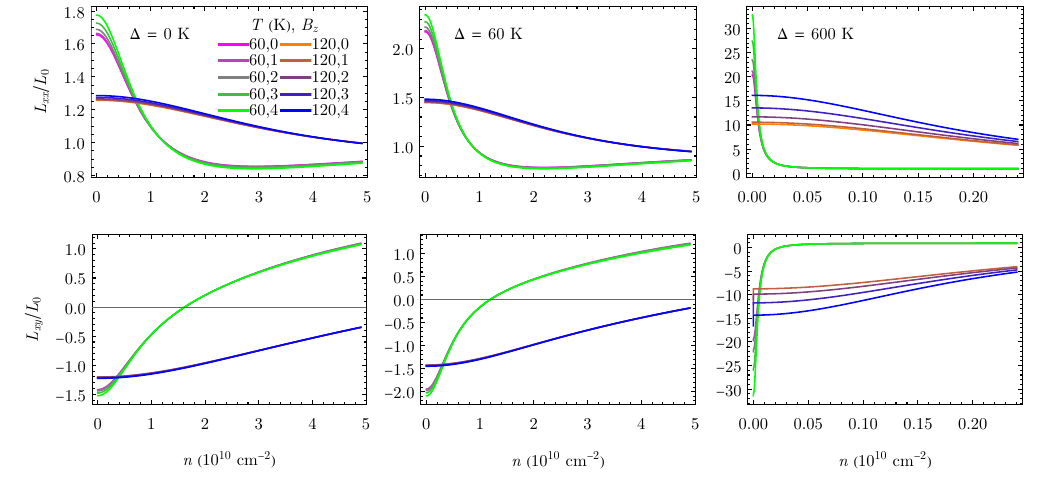}
  \caption{
    Longitudinal and transverse Lorenz ratio for MLG as a function of $n$, for various choices of $\Delta$, $T$, and $B_z$.
    The scattering parameters and the unit of $B_z$ are the same as in Fig.~\ref{fig:MLG_T}.
  }
  \label{fig:MLG_n}
\end{figure*}

The MLG is typically modeled by linearly dispersive gapless conduction and valance bands~\cite{DasSarma2011}.
However, we consider the possibility of a gap opening as in Ref.~\cite{Tu2023}, which may be due to the hBN substrate~\cite{DeanPrivate,Hunt2013,RibeiroPalau2018,Finney2019,Yankowitz2018}.
Since the exact behavior near the gap is unknown, we use the simplest model for the gap as in Ref.~\cite{Tu2023} (note that the result for parabolic dispersion is qualitatively similar as shown in Sec.~\ref{sec:BLG}, so the exact shape near the gap should not affect our qualitative result),
\begin{align}\label{eq:dispersion}
  \varepsilon_{+}(\mathbf{p})&=+ v_F |\mathbf{p}|\nonumber\\
  \varepsilon_{-}(\mathbf{p})&=-v_F |\mathbf{p}|-\Delta\,,
\end{align}
where $\Delta$ is the size of the gap and $v_F\sim 1\times10^6\,\mathrm{m/s}$ is the Fermi velocity of graphene. The subscripts label the conduction ($+$) and the valance ($-$) band.
The density of states is (including the spin degeneracy $g_s=2$ and valley degeneracy $g_v=2$)
\begin{align}
  D_+(\varepsilon)&=\frac{2\varepsilon}{\pi\hbar^2 v_F^2}\quad\text{for }\varepsilon>0\nonumber\\
  D_-(\varepsilon)&=\frac{2(-\Delta-\varepsilon)}{\pi\hbar^2 v_F^2}\quad\text{for }\varepsilon<-\Delta\,.
\end{align}

For the relaxation time $\tau$, it is known that the dominant transport mechanisms in graphene are the scattering by
 short-range disorder, long-range disorder, and acoustic phonon~\cite{DasSarma2011}.
 As in Ref.~\cite{Tu2023}, we consider only these three mechanisms, using the phenomenological model derived from Refs.~\cite{Hwang2008a,Hwang2009}:
\begin{equation}
  \tau_+(\varepsilon)=\frac{1}{A\varepsilon+BT\varepsilon+\frac{C}{\varepsilon}},\quad \tau_-(\varepsilon)=\tau_+(-\Delta-\varepsilon).
\end{equation}
Here the parameters $A$, $B$, $C$ represent the scattering strengths of short-range disorder, acoustic phonon, and long-range Coulomb disorder, respectively (the magnetic field is denoted by ``$B_z$'' to avoid confusion with the coefficient ``$B$'' here).
At zero magnetic field, only the ratios between the parameters affect the Lorenz ratio.
Although scaling $\tau$ by a constant affects the Lorenz ratio at nonzero $B_z$, it only changes the unit of it.
We try several different combinations of the parameters (as in Ref.~\cite{Tu2023}), finding that they give the same qualitative results.
Since the experimental value of these parameters is unknown, we will just choose one set of typical $(A/C,B/C)$ for each $\Delta$, and present the results using a unit of $B_z$ that depends on $C$.
This also means that the maximum $B_z$ that satisfies the weak requirements cannot be pinned down in our results, as the actual value depends on $C$, which is unknown.
With such a large number of unknown parameters in the problem, our goal is neither data fitting nor precise quantitative predictions, but aiming at the expected qualitative dependence of the effective Lorenz ratio in the presence of a finite magnetic field.

We present the magnetic-field-dependent result of the Lorenz ratio as a function of $T$ in Figure~\ref{fig:MLG_T} and as a function of $n$ in Figure~\ref{fig:MLG_n}.
We observe that (1) the Wiedemann-Franz law is asymptotically satisfied for both the longitudinal and transverse component as $T\to 0$; (2) for the longitudinal component, the finite temperature peak is enhanced by the magnetic field, and the enhancement is larger at lower density; (3) the position of the finite temperature peak is almost independent of $B_z$; (4) for the transverse component, the value can be either positive or negative, depending on the parameter regime, which is expected due to the complex behavior or the electron and holes in the presence of the magnetic field.
For completeness, we also present the Lorenz ratio as a function of $B_z$ for a particular choice of parameters in the left column of Fig.~\ref{fig:Bz}.
We caution however that our theory would not apply at ``larger'' values of $B_z$ where strong field effects such as Landau quantization would come into play.

\section{Bilayer graphene}\label{sec:BLG}

\begin{figure*}
  \includegraphics[trim=0 0 0 0, clip]{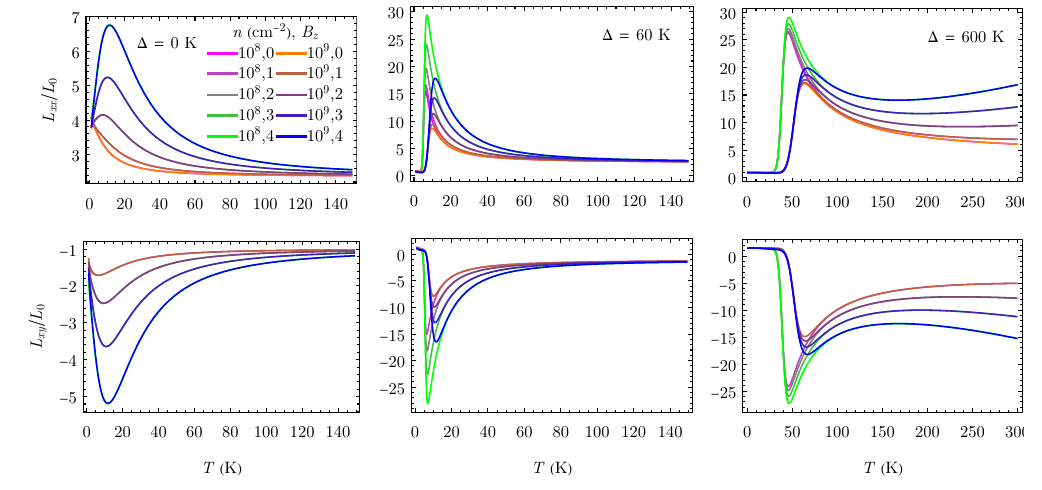}
  \caption{
    Longitudinal and transverse Lorenz ratio for BLG as a function of $T$, for various choices of $\Delta$, $n$, and $B_z$.
    The scattering parameters in the cases of $\Delta=0\,\mathrm{K}$ (first column) and $\Delta=60\,\mathrm{K}$ (second column) are $A/C=0.0167\,\mathrm{K}^{-1}$, $B/C=2.78\times10^{-3}\,\mathrm{K}^{-2}$, and that in the case of $\Delta=600\,\mathrm{K}$ (third column) are $A/C=1.67\times10^{-6}\,\mathrm{K}^{-1}$, $B/C=2.78\times10^{-9}\,\mathrm{K}^{-2}$.
    The unit of $B_z$ in the cases of $\Delta=0,60\,\mathrm{K}$ is $80\,Ce^{-1}v_F^{-2}$ and in the case of $\Delta=600\,\mathrm{K}$ is $2\,Ce^{-1}v_F^{-2}$.
  }
  \label{fig:BLG_T}
\end{figure*}

\begin{figure*}
  \includegraphics[trim=0 0 0 0, clip]{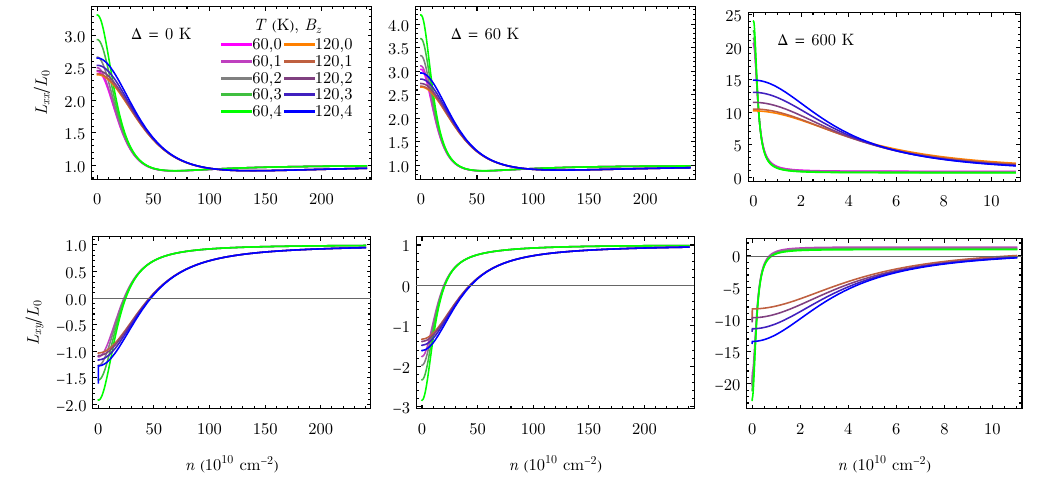}
  \caption{
    Longitudinal and transverse Lorenz ratio for MLG as a function of $n$, for various choices of $\Delta$, $T$, and $B_z$.
    The scattering parameters and the unit of $B_z$ are the same as in Fig.~\ref{fig:BLG_T}.
  }
  \label{fig:BLG_n}
\end{figure*}

\begin{figure}
  \includegraphics[trim=22 0 0 0, clip]{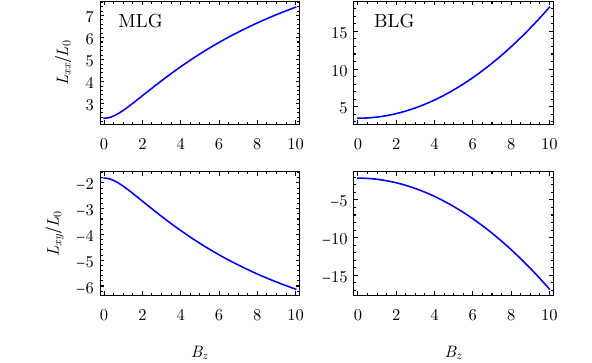}
  \caption{
    Longitudinal and transverse Lorenz ratio for MLG (left column) and BLG (right column) as a function of $B_z$ at $\Delta=60\,\mathrm{K}$, $T=40\,\mathrm{K}$, $n=10^9\,\mathrm{cm}^{-2}$.
    The unit of $B_z$ is $Ce^{-1}v_F^{-2}$ for MLG and $80\,Ce^{-1}v_F^{-2}$ for BLG.
    The scattering parameters are the same as in the corresponding cases in Figs.~\ref{fig:MLG_T} and \ref{fig:BLG_T}.
  }
  \label{fig:Bz}
\end{figure}

Near the Fermi surface, the BLG is modeled by parabolic dispersive conduction and valance bands~\cite{DasSarma2011}. As in the case of MLG, we consider the situation where a gap is opened, and use the simplest model:
\begin{align}
  \varepsilon_{+}(\mathbf{p})&=+ \frac{|\mathbf{p}|^2}{2m}\nonumber\\
  \varepsilon_{-}(\mathbf{p})&=- \frac{|\mathbf{p}|^2}{2m}-\Delta\,,
\end{align}
where $\Delta$ is the size of the gap and $m\approx 0.2\,\mathrm{eV}/v_F^2$ is the effective mass~\cite{DasSarma2011}.
The density of states is (including the spin degeneracy $g_s=2$ and valley degeneracy $g_v=2$):
\begin{equation}
  D_\pm(\varepsilon)=\frac{2m}{\pi\hbar^2}\quad\text{for }\varepsilon>0\nonumber\text{ or }\varepsilon<-\Delta.\\
\end{equation}

We use the same scattering mechanisms for $\tau$ as in MLG, but the scattering exponents are differentbecause of the modified band structures. From the result of Ref.~\cite{DasSarma2010}, we use the following phenomenological model:
\begin{equation}
  \tau_+(\varepsilon)=\frac{1}{A+BT+\frac{C}{\varepsilon}},\quad \tau_-(\varepsilon)=\tau_+(-\Delta-\varepsilon).
\end{equation}
Here the parameters $A$, $B$, $C$ represent the scattering strengths of short-range disorder, acoustic phonon, and long-range Coulomb disorder, respectively, as in the MLG case.

We present the result of the Lorenz ratio as a function of $T$ in Figure~\ref{fig:BLG_T} and as a function of $n$ in Figure~\ref{fig:BLG_n}.
We observe that the behavior is qualitatively similar to the case of MLG.
In particular, there is a high finite-temperature peak when there is a gap but only manifests a small peak when there is no gap, as in the case of MLG found in Ref.~\cite{Tu2023}, and the peak becomes higher for nonzero $B_z$.
However, the quantitative details are different from that of MLG (note that different combinations of scattering parameters can also lead to some difference in the quantitative details, so one should not compare the MLG and BLG results presented here too literally).
Again, for completeness, we present the Lorenz ratio as a function of $B_z$ for a particular choice of parameters in the right column of Fig.~\ref{fig:Bz}.

\section{conclusion}\label{sec:conclusion}
Using the Boltzmann transport theory with a magnetic field, we show that the large finite-temperature peak of $L_{xx}/L_0$, observed in Ref.~\cite{Crossno2016} and possibly (qualitatively) explained by our previous paper~\cite{Tu2023}, is enhanced, but not shifted much, by the presence of the magnetic field.
In addition, we note that the sign of $L_{xy}/L_0$ may either be positive or negative, depending on the parameter regime.
Such qualitative behaviors are the same in both MLG and BLG.

We emphasize that our current work, although a generalization of our earlier work~\cite{Tu2023}, is both important and novel by virtue of the singular importance of the Crossno {\it et al.\ }experiment~\cite{Crossno2016} we are investigating.
The original experiment was interpreted as a breakdown of the Fermi liquid theory in graphene, and as such had a huge impact.
We pointed out in~\cite{Tu2023} that there is a compelling Fermi liquid interpretation of the experiment~\cite{Crossno2016}, and unpublished data appear to support our claim that the maximum value of $L/L_0$ should be $\sim 4$ unless a gap opens up at the Dirac point due to the hBN substrate.
The experimentalists pointed out to us that one novel way to explore our predictions and understand the correct physics would be to apply a magnetic field and study the magnetic field dependence of $L/L_0$, which provides a new tuning parameter to study the system definitively~\cite{KimPrivate}.
Our current work provides definitive and comprehensive predictions for how $L/L_0$ should behave as a function of an applied magnetic field, enabling a decisive experimental settlement of the Fermi liquid versus non-Fermi liquid question once and for all by varying an applied magnetic field in the experiment and comparing with the results presented in the current work.
We emphasize that magnetotransport brings in qualitatively new physics by introducing chirality and Hall effect into the problem, and the system now has both longitudinal and transverse responses, introducing a profound new direction to be tested experimentally~\cite{KimPrivate}.
Such magnetic-field-dependent experiments are currently underway, and our predictions (before the fact) obviously are important and necessary to settle what is certainly an important conceptual question:  Is doped graphene a Fermi liquid or not?  Our magnetic-field-dependent results when compared with the ongoing experimental efforts should answer this profound question in the near future.

Note that we do not claim that our previous paper \cite{Tu2023} unambiguously explained the observation in Ref.~\cite{Crossno2016}, and hence do not claim that the results in this paper necessarily correspond to the reality if one adds a magnetic field to the experimental setup in that paper.
In particular, we are not claiming that the simple assumptions used in these two papers accurately correspond to the experimental system, nor that the simplfied theory can give all the quantitative details that agree with experimental data.
Our purpose is to provide a possible explanation, as all the previously attempted explanations are not very successful.
We are successful to the extent that the addition of a single parameter, namely a gap, can provide an explanation for the intriguing data of Ref.~\cite{Crossno2016}. Now, we develop the same theory in the presence of a magnetic field, providing further motivation for more experiments to clarify the physics of the MLG and BLG Wiedemann-Franz law.

If future experimental results with the addition of a weak magnetic field agree qualitatively with the results here, then one may say that our explanation \cite{Tu2023} is likely correct (of course, more experiments, such as deliberately inducing a gap, may also be necessary to settle down the explanation~\footnote{
  Another important observable for the experiments is the Hall angle $\cot \theta_H=\sigma_{xx}/\sigma_{xy}$, which can be calculated directly from the first two lines of Eq.~(\ref{eq:transport}). In particular, for the phonon-dominated regime of the gapless model of MLG, we have $\cot\theta_H\sim T$ when $T\ll T_F$ and $\cot\theta_H\sim T^5/\log T$ when $T\gg T_F$.
}). In this case, one may then try to extract the parameters from the data, and establish a more quantitative microscopic theory for the transport in MLG as well as BLG.
On the other hand, if future experimental results disagree qualitatively with the results here, then it would imply that the finite temperature peak observed in~\cite{Crossno2016} cannot be explained just by considering bipolar diffusion and the induced gap.
In that case, more theoretical works would be necessary to solve the puzzle presented in~\cite{Crossno2016}.
Such experiments to understand the temperature dependence of the Wiedemann-Franz law in graphene in the presence of a magnetic field is currently ongoing~\cite{KimPrivate}, and hopefully, we will have a resolution of the puzzle posed by Ref.~\cite{Crossno2016} in the near future.

\section*{Acknowledgment}

This work is supported by the Laboratory for Physical Sciences.
 
\bibliographystyle{apsrev4-2}
\bibliography{references}

\end{document}